\documentclass[aps,prl,twocolumn,showpacs,groupedaddress]{revtex4}
\usepackage{graphicx}  
\usepackage{dcolumn}   
\usepackage{bm}        
\usepackage[english]{babel}
\usepackage{amsfonts,amsmath,amssymb,mathrsfs}
\usepackage{epstopdf}
\usepackage{subfigure}
\usepackage{color}

\def\a{\alpha}
\def\r{\rho}
\def\s{\sigma}
\def\t{\tau}
\def\m{\mu}
\def\n{\nu}
\def\k{\kappa}
\def\th{\theta}
\def\g{\gamma}\def\G{\Gamma}
\def\L{\Lambda}\def\l{\lambda}
\def\D{\Delta}
\def\la{\langle}
\def\ra{\rangle}
\def\o{\omega}\def\O{\Omega}
\def\d{\delta}
\def\p{\partial}

\def\half{\textstyle{\frac{1}{2}}}

\def\bdoc{\begin{document}}
\def\edoc{\end{document}}

\def\beq{\begin{equation}}
\def\eeq{\end{equation}}
\def\bea{\begin{eqnarray}}
\def\eea{\end{eqnarray}}
\def\ben{\begin{enumerate}}
\def\een{\end{enumerate}}
\def\la{\langle}
\def\ra{\rangle}
\def\a{\alpha}
\def\b{\beta}
\def\g{\gamma}\def\G{\Gamma}
\def\d{\delta}\def\D{\Delta}
\def\e{\epsilon}
\def\z{\zeta}

\def\th{\theta}
\def\k{\kappa}
\def\l{\lambda}
\def\m{\mu}
\def\n{\nu}
\def\o{\omega}
\def\p{\pi}
\def\r{\rho}
\def\s{\sigma}
\def\t{\tau}
\def\L{{\cal L}}
\def\S{\Sigma }
\def\gsim{\; \raisebox{-.8ex}{$\stackrel{\textstyle >}{\sim}$}\;}
\def\lsim{\; \raisebox{-.8ex}{$\stackrel{\textstyle <}{\sim}$}\;}
\def\gtrsim{\gsim}
\def\lessim{\lsim}
\def\loc{{\rm local}}
\def\vm{v_{\rm max}}
\def\bh{\bar{h}}
\def\del{\partial}
\def\nab{\nabla}
\def\half{{\textstyle{\frac{1}{2}}}}
\def\fourth{{\textstyle{\frac{1}{4}}}}

\def\bD{{\bf D}}
\def\bE{{\bf E}}
\def\bF{{\bf F}}
\def\bB{{\bf B}}
\def\bP{{\bf P}}
\def\bV{{\bf v}}
\def\bv{{\bf v}}
\def\bx{{\bf x}}
\def\by{{\bf y}}
\def\bz{{\bf z}}
\def\ba{{\bf a}}
\def\bd{{\bf d}}
\def\bs{{\bf s}}
\def\bn{{\bf n}}
\def\bp{{\bf p}}

\def\O{\Omega}

\def\br{{\bf r}}
\def\bnab{{\bf \nab}}

\def\tE{\tilde{E}}
\def\tL{\tilde{L}}
\def\Horava{Ho\v{r}ava }

\begin{document}

\title{Extended \Horava gravity and
Einstein-aether theory}
\author{Ted Jacobson}
\affiliation{Center for Fundamental Physics,  University of Maryland, College Park, MD 20742-4111, USA}
\date{\today} 
\begin{abstract}
Einstein-aether theory is general relativity coupled to a
dynamical, unit timelike vector. If this vector is restricted
in the action to be hypersurface orthogonal, 
the theory is identical to the IR limit
of 
the extension of \Horava gravity proposed
by Blas, Pujol\`{a}s and Sibiryakov.
Hypersurface orthogonal solutions of 
Einstein-aether theory are solutions to the IR limit of 
this theory, 
hence numerous results
already obtained for Einstein-aether theory carry
over.
\end{abstract}  
\maketitle

Much interest has recently been focused on 
Ho\v{r}ava-Lifshitz gravity \cite{Horava:2009uw}, which proposes
the possibility of a renormalizable, non-Lorentz-invariant 
UV completion of general relativity. 
There are so-called projectable 
and nonprojectable versions of this proposal, and both
have been shown to suffer from various problems 
(instabilities, overconstrained evolution, or strong coupling at low energies)
related to a badly behaved scalar mode of gravity
brought on by the presence of a nondynamical spatial foliation 
in the action  \cite{Cai:2009dx}. 
A proposal for evading all of these problems,
put forth by Blas, Pujol\`{a}s and Sibiryakov (BPS) \cite{Blas:2009qj},
is an ``extension" of Ho\v{r}ava gravity.
I will call it here BPSH gravity, and below $T$-theory
for reasons to become clear. 
One can view this extension
as promoting the fixed foliation to a dynamical one.
This extension could still possess strong coupling at low energy 
\cite{Papazoglou:2009fj}, but it is also possible that higher derivative
terms in the action become important below the strong coupling
energy scale and prevent this \cite{Blas:2009ck}.

It was remarked in Ref.\ \cite{Blas:2009ck}
that this extended \Horava theory is related to a restricted version of 
Einstein-aether theory (see also Ref.\ \cite{Germani:2009yt}
for the nonextended case), which is general relativity coupled to a 
dynamical unit timelike vector field (for  recent review see 
Ref.\ \cite{Jacobson:2008aj}).
The restriction amounts to assuming the vector field
is hypersurface orthogonal.
The purpose of this article is to clarify the relation between these
two theories, both at the level of the defining action principles 
and at the level of solutions to the equations of motion. 
In particular, the lowest dimension
terms (the IR limit) of the BPSH gravity action are equivalent 
to those of Einstein-aether theory, 
when the aether vector is assumed to be 
hypersurface orthogonal. It will be shown that any 
hypersurface orthogonal solution to Einstein-aether theory
is a solution to the IR limit of 
BPSH gravity, although the converse does
not appear to be true. In particular, since all spherically symmetric 
aether fields are hypersurface orthogonal, 
the spherically symmetric vacuum, star, black hole, and collapsing
star solutions,
and FRW cosmological solutions
to Einstein-aether theory are 
all solutions to BPSH gravity. Moreover, some results about the coupling
constants and PPN parameters of Einstein-aether theory 
carry over.

BPSH gravity has been formulated as a theory
with a preferred spacetime foliation, defined by a 
time coordinate $t$ and space coordinates 
$x^i$. The spacetime metric is given by 
\beq\label{ADM}
ds^2 = N^2 dt^2 - h_{ij}(dx^i - N^i dt)(dx^j - N^j dt),
\eeq
where $N$ and $N^i$ are the lapse function and 
shift vector.
The action for the IR sector of BPSH theory is
\beq\label{SBPSH}
S= \frac{1}{16\pi G_{H}}\int dt d^3x \, N\sqrt{h}(K_{ij}K^{ij} - \l K^2 
+ \xi {}^{(3)}\!R + \a a_ia^i).
\eeq
Here spatial indices are raised and lowered using $h_{ij}$, 
$K_{ij}=(\dot{h}_{ij} + D_iN_j + D_j N_i)/2N$ is the extrinsic curvature 
of 
the spatial surface, dot denotes derivative with respect to $t$, 
$D_i$ is the spatial covariant derivative, 
$K=h^{ij}K_{ij}$, ${}^{(3)}\!R$ is the spatial Ricci curvature scalar,
and $a_i = -\partial_i\ln N$. The coefficients $\l$, $\xi$, and 
$\a$ are dimensionless constants, and $G_H$ is what 
BPS denote by $M_P^2/2$.

It is helpful to understand this BPSH gravity 
theory using a covariant formalism. 
I shall start from scratch defining a theory 
motivated by symmetry principles, and then 
show that the resulting theory is equivalent to 
BPSH gravity. I focus here purely on the IR limit.

Suppose we wish to write down a generally covariant
theory that depends on a spacetime metric, but also
on a foliation of
spacetime by spacelike surfaces. These surfaces 
define a notion of ``cosmic simultaneity," quite 
alien to the spirit of general relativity. However, 
given various difficulties constructing a viable
theory of quantum gravity (including nonrenormalizability
and UV completion, and the problem of time), 
it is interesting to ask what such a theory would look like
and whether it could solve some or all of these
problems and be phenomenologically viable.

A spacelike foliation of spacetime  
can be defined by the level sets of a suitably
behaved scalar function $T$. If this function is
a dynamical variable in the theory, then general
covariance is preserved. 
A monotonically related function 
defines the same foliation. If the 
theory is to not depend on the time labeling, but rather 
only on the choice of surfaces, the action should depend 
on $T$ only via the unit (co)vector 
\begin{equation}\label{u}
u_a=W T_{,a}, \quad {\rm with}\quad  W=(g^{ab}T_{,a}T_{,b})^{-1/2},
\end{equation}
where the subscript ``${,a}$" denotes the gradient.
The convention used here for metric
signature is $({+}{-}{-}{-})$.
This vector is {\it hypersurface orthogonal}, 
i.e. $u_a v^a=0$ for any vector $v^a$ 
tangent to a surface of constant $T$. 
Conversely, any hypersurface orthogonal
vector can be written in the form   $WT_{,a}$ 
for some pair of scalar functions $T$ and $W$.

To formulate a local, 
generally covariant dynamical theory of the spacetime metric and
the $T$ function via the $u_a$ vector, we 
specify a Lagrangian scalar density. The classification 
of spacetime scalars that can be written
using $g_{ab}$ and $u_a$, with up to two
derivatives,  was already considered 
\cite{Jacobson:2000xp,Jacobson:2008aj}
in the context of 
Einstein-aether theory, or {\it ae-theory} for short. 
The only difference between 
ae-theory and what I will call {\it T-theory} for short,
is that in ae-theory the 
aether, i.e. the unit vector field, is not required 
to be hypersurface orthogonal.
(Actually the aether has usually been taken to
be contravariant rather than covariant, 
but this changes only appearances.)
This means that in ae-theory the aether has 
three degrees of freedom at each 
spacetime point, whereas in the present case 
it has just one, coming from the 
choice of the time function $T$.

Up to total derivative terms, the most general
action for ae-theory (aside from matter couplings) is
\beq \label{S}
S = \frac{1}{16\pi G_{\ae}}\int \sqrt{-g}~ (-R + L_{\ae})
~d^{4}x \eeq
where $R$ is the 4D Ricci scalar and
\beq \label{Lae}
L_{\rm ae} = -M^{abmn} \nabla_a u_m \nabla_b u_n,
\eeq
with $M^{abmn}$ is defined as
\beq M^{abmn} = c_1 g^{ab}g^{mn}+c_2g^{am}g^{bn}
+c_3 g^{an}g^{bm}+c_4 u^a u^b g_{mn}. \eeq
The $c_i$ are dimensionless coupling constants,
and it is assumed that $u_m$ is constrained to be a unit 
vector, $g^{mn}u_m u_n=1$. 
Note that
since the covariant derivative operator $\nabla_a$ involves
derivatives of the metric through the connection components, and
since the unit vector is nowhere vanishing, the terms quadratic in
$\nabla u$ also modify the kinetic terms for the metric.

To pass to  $T$-theory we just substitute (\ref{u}) 
for $u_m$ in the Lagrangian (\ref{Lae}).
The equations of motion are the conditions that the
action be stationary under variation of the metric and 
the scalar function $T$.
The resulting action looks dangerous because it has two 
explicit derivatives,
and $u_a=WT_{,a}$ already has one implicit derivative. 
Hence it appears that the equations of motion for $T$ 
will have fourth derivatives,
which does not sound healthy. However, 
since the theory is generally covariant,
we may always express the field equations using $T$ itself
as one of the spacetime coordinates. Then we have 
$u_a=\delta_{aT}(g^{TT})^{-1/2}$, which contains no derivatives.
Still the variation of $T$ in the action will produce an
equation of motion that is third order in derivatives.
However, as we shall see it remains second order in time
derivatives. 

Another issue is the timelike character
of $T_{,a}$. It is not clear whether the dynamics of the theory
somehow manages to preserve this condition.
If not, then where $T_{,a}$ becomes null the unit vector
$u_a$
will diverge, and most likely the metric would become 
singular as well. It seems an important question to determine 
whether such singularities can arise ``unprovoked", and/or more
visibly than those hidden by black hole horizons in general 
relativity.

An important observation we can now make is that 
{\it the $T$ field equation is 
implied by the Einstein equation}. One way to think about this
is to consider ordinary Einstein gravity coupled to a scalar
field. Because of the Bianchi identity $\nab^a G_{ab}=0$, 
the Einstein equation implies conservation 
of the stress energy tensor. For a scalar field this
is enough to imply the matter field equation, unless the
scalar field is constant. Another way to see this is to 
appeal directly to diffeomorphism invariance of the action. 
Suppose we couple $T$-theory to generic matter fields
denoted $\psi$, so the action is schematically $S[g,\psi,T]$.
Under a diffeomorphism generated by a vector field $\xi$ the
action is invariant, so we have 
\bea
\d_\xi S &=&\int (\d S/\d g){\cal L}_\xi g + (\d S/\d \psi) {\cal L}_\xi \psi + 
(\d S/\d T){\cal L}_\xi T \nonumber\\
&=&0,
\eea
where ${\cal L}_\xi$ is the 
Lie derivative. Now suppose the Einstein equation is
satisfied, so that $\d S/\d g=0$, and suppose further that
the matter field equations are satisfied, so that 
$\d S/\d \psi=0$. Then at such solutions we have the 
identity $\int (\d S/\d T){\cal L}_\xi T=0$ for all vector fields
$\xi^a$. Unless
$T$ is constant --- which it cannot be if it is to define a foliation --- the Lie derivative ${\cal L}_\xi T$ 
can be freely varied at a point by varying the choice of
$\xi^a$, hence it must be that $\d S/\d T=0$, i.e. the
$T$ field equation is satisfied.

Since the $T$ field equation need not be imposed explicitly,
we can adopt $T$ as one of the spacetime coordinates,
and write the theory in this ``$T$-gauge" {\it before} varying the
remaining fields, using the
3+1 decomposition of the metric in (\ref{ADM}).
Then since $u=W dT$ is a unit 1-form, the function $W$
is evidently the same as the lapse $N$.

Decomposing the volume element yields $\sqrt{-g}=N\sqrt{h}$.
The covariant derivative of $u_m$ decomposes as
\beq 
\nab_a u_b = - K_{ab} -u_a a_b,
\eeq
 where $K_{ab}$ is the extrinsic curvature of the surfaces
 normal to $u_a$, and $a_b$ is the acceleration of the 
 normal curves. The extrinsic curvature is symmetric,
 and it and the acceleration are both  
 spatial  ($K_{ab} u^b=0$ and $a_b u^b=0$).
 Using the spatial coordinates $x^i$ the acceleration
 can be expressed as $a_i = -(\ln N)_{,i}$.
 The aether Lagrangian 
 (\ref{Lae}) can therefore be expressed in the form 
\beq \label{Laes}
L_{\rm ae} = -c_{13}K_{ij}K^{ij}-c_2 K^2 +c_{14}a_ia^i,
\eeq
where, $c_{13}=c_1+c_3$, $c_{14}=c_1+c_4$, and,
as in (\ref{SBPSH}), the spatial indices are raised
with $h^{ij}$. 
The 3+1 decomposition of the $-R$ in the 
Einstein-Hilbert action adds
to the Lagrangian 
$K_{ij}K^{ij}- K^2 + {}^{(3)}\!R$. 

We conclude that the $T$-theory action
(\ref{S}) and the BPSH action (\ref{SBPSH})
are identical, with the following relations between the 
various coefficients: 
\beq
G_H/G_{\ae}=\xi= 1/(1-c_{13}),\quad  \a/\xi = c_{14},\quad 
\l/\xi =  1+c_2.
\eeq
In Ref.\ \cite{Blas:2009qj},  $\xi$ is fixed to unity by 
choosing the scale of the $t$ coordinate.
Once matter is present, that rescaling
is no longer available if matter is to couple minimally 
to the spacetime metric. 

Note that athough we begin with four 
coefficients $c_{1,2,3,4}$
in the most general $T$-theory action, when 
expressed in 3+1 form in the $T$-gauge only
three independent combinations of these parameters 
enter (\ref{Laes}). From the covariant 
point of view this happens because, when 
$u_a$ is hypersurface orthogonal, there is a
relation between three of the terms in the Lagrangian
(\ref{Lae}). This can be seen by considering
the twist 3-form, $\omega=u\wedge du$, which vanishes identically  when 
$u=NdT$. In terms of the dual vector 
$\omega^a = \epsilon^{abcd}u_b\nabla_c u_d$
we have the identity
\beq\label{twist}
 \omega_a \omega^a = 
 -2(\nabla_{a} u_b)(\nabla^{[a} u^{b]}) 
 +(u^b \nabla_b u_a)(u^c \nabla_c u^a),\eeq
where the square brackets denote index antisymmetrization.
When $u_a$ is hypersurface orthogonal $\omega^a$ vanishes, 
so the $c_1$, the $c_3$ or the
$c_4$ term in the Lagrangian can be written in terms of the other two.
Alternatively, one can use this identity to 
remove the square of the 
antisymmetric part $(\nab_{[a} u_{b]}) (\nab^{[a} u^{b]})$,
replacing both the $c_1$ and $c_3$ terms by a single term
of the form $(\nab_{(a} u_{b)}) (\nab^{(a} u^{b)})$
(where the round brackets denote index symmetrization).

One more term may be eliminated from the action (\ref{S}) 
by making a field redefinition of the 
metric \cite{BarberoG.:2003qm,Foster:2005ec}
\beq
g'_{ab} = g_{ab}+(\z-1) u_a u_b,
\eeq
which ``stretches" the metric tensor in the aether
direction by a positive factor $\z$. 
(A negative factor would return a Euclidean signature 
metric \footnote{Such field redefinitions, with 
a hypersurface orthogonal vector field, were introduced
in Ref.\ \cite{BarberoG.:1995ud} 
as a way to relate solutions to the Einstein equation
with Lorentzian and Euclidean signature.}.)
This does not change the function $T$, but the 
unit vector does change, to $u'_a = \sqrt{\z}u_a$.
If matter is coupled minimally to the
metric $g_{ab}$, then it is not coupled minimally
to $g'_{ab}$ but rather to $g'_{ab}-(1-1/\z) u'_a u'_b$.
However, for considerations not involving matter, such a field 
redefinition can simplify the theory. 
The action (\ref{S})
for ($g'_{ab}$, $u'_a = \sqrt{\z}u_a$) takes the same
form as that for ($g_{ab}$, $u_{a}$), with new
coefficients $c'_i$. The relation between the $c'_i$
and $c_i$ was worked out in Ref.\ \cite{Foster:2005ec}, and
is conveniently given in terms of certain combinations
with simple scaling behavior:
\bea
c'_{14} &=& c_{14}\nonumber\\
c'_{123} &=& \z c_{123}\nonumber\\
c'_{+}-1 &=& \z (c_{+}-1)\nonumber\\
c'_{-}-1 &=& \z^{-1}(c_{-}-1), \label{redef} \eea
where 
$c_{123}=c_1+c_2+c_3$ and 
$c_{\pm}=c_1\pm c_3$. 
For example, 
one can arrange for $c'_{+}=0$ by choosing
$\z=1/(1-c_{+})$ (provided $c_{+}<1$).
Thus by a metric redefinition the symmetrized term
$(\nab_{(a} u_{b)}) (\nab^{(a} u^{b)})$
can be eliminated from the Lagrangian.
Then, using the
twist identity (\ref{twist}), one can either replace the
remaining, antisymmetrized term by an acceleration squared term,
or vice versa. The first way yields the Lagrangian 
\beq L_{\rm ae} = 
\g_2 (\nabla_a u^a)^2 + \g_4 (u^b \nabla_b u_a)(u^c \nabla_c u^a),
\eeq
where $\g_{2,4}$ depend on the original values
$c_{1,2,3,4}$. The second way yields the equivalent Lagrangian
\beq L_{\rm ae} =
\g_2 (\nabla_a u^a)^2+\g_-\nabla_{[a} u_{m]}\nabla^{[a} u^{m]} .
\eeq

The equations of motion of $T$-theory are closely related to those
of ae-theory. Let $E^n$ denote the variation of the action with 
respect to $u_n$,
\beq
E^n = 16\pi G_{\ae}\, \frac{\d S}{\d u_n}.
\eeq
The variations of $u_n$ induced by 
variations of $T$ and of the inverse metric $g^{ab}$ are 
\bea
\d_T u_n &=& N(\d_n^m - u_nu^m)\nab_m\d T\\
\d_g u_n &=& -\half u_n u_a u_b \d g^{ab}.\label{dudg}
\eea
The $T$ equation of motion is thus 
\beq\label{Tfe}
\nab_m\Bigl(N(\d^m_n - u^m u_n)E^n\Bigr)=0.
\eeq
As mentioned earlier, if $T$ is used as one of the
coordinates, then $E^n$ has just two derivatives in
it, and then the $T$ field equation (\ref{Tfe})
is of third order in derivatives. However, 
due to the presence of the spatial
projection $\d^m_n - u^m u_n$, the derivative in
the outer $\nab_m$ operator is purely spatial, so the
equation is of only second order in derivatives with respect to $T$.

The $g^{ab}$ field equation, including matter fields, is
\beq\label{gfe}
G_{ab} = T^{\ae}_{ab} + 8\pi G_{\ae} T^{\rm matter}
\eeq
where $G_{ab}=R_{ab}-\half R g_{ab}$ 
is the usual Einstein tensor and $T^{\ae}_{ab}$ denotes the
aether stress tensor, which includes the contributions from 
the variations in the aether action (\ref{S}) 
of the explicit metrics in $\sqrt{-g}$ and 
$M^{abmn}$, 
the metrics in the covariant derivative operators, and those
buried in the definition of $u_m$. In particular,
according to (\ref{dudg}) the latter contribute
\beq\label{uuterm}
-\half (E^n u_n) u_a u_b
\eeq
to $T^{\ae}_{ab}$, which has the form of the stress tensor of
a pressureless dust.

Let us now compare the form of the field equations
(\ref{Tfe}) and (\ref{gfe}) to those of ae-theory. 
In ae-theory, the fundamental field $u_m$ is
constrained to satisfy $g^{mn}u_mu_n=1$. 
This constraint has
usually been implemented with a Lagrange multiplier term
in the action, but it can also be taken into account by 
restricting the variations to satisfy 
$\d g^{mn}u_mu_n+ 2 u^n\d u_n=0$.
Thus the part of the aether variation 
orthogonal to $u^n$ is unconstrained, while the 
part parallel to $u_n$ is given by 
$\d_{\parallel} u_n = -\half u_n u_a u_b \d g^{ab}$, 
just as in (\ref{dudg}). Thus the metric variation,
accompanied by this parallel aether variation, 
yield the Einstein equation (\ref{gfe}) as before, with 
the contribution (\ref{uuterm}) in the aether stress tensor.
The orthogonal aether variation yields
\beq\label{ufe}
(\d^m_n - u^m u_n)E^n=0.
\eeq
This implies the $T$-theory field equation (\ref{Tfe}),
although the converse is not true. 
We conclude that {\it any hypersurface orthogonal
solution to Einstein-aether theory is also a solution to 
$T$-theory, i.e.\ to BPSH gravity}. 

In particular, since 
any spherically symmetric aether field is hypersurface orthogonal,
any 
such ae-theory solution is a 
$T$-theory solution. An example is the vacuum solution
to $T$-theory found recently by Kiritsis  \cite{Kiritsis:2009vz}
which is identical to the one found previously for ae-theory \cite{Eling:2006df}. 
A second, non asymptotically flat solution of $T$-theory 
was found in Ref.\ \cite{Kiritsis:2009vz}. This is also an
ae-theory solution: it corresponds to the interior part of the 
wormhole in the global solution found in Ref.\ \cite{Eling:2006df}.
In fact, for an aether parallel to the time translation Killing field
in spherical symmetry, the aether equation is automatically
satisfied \cite{Eling:2006df}, hence {\it all} such $T$-theory solutions
are ae-theory solutions. More generally, however, it appears that
not all spherically symmetric $T$-theory solutions need be ae-theory solutions.

The conclusion that spherical ae-theory solutions
are $T$-theory  solutions applies also to the 
black hole solutions in Ref.\ \cite{Eling:2006ec},
the neutron star solutions in Ref.\ \cite{Eling:2007xh},
and even the time-dependent collapse solutions
of Ref.\ \cite{Garfinkle:2007bk}.
It also applies to the homogeneous isotropic 
cosmological solutions
discussed in Refs. \cite{Mattingly:2001yd,Carroll:2004ai}.
Moreover, we can infer that
the relation \cite{Carroll:2004ai} 
between Newton's constant $G_N$
and the coefficient $G_{\ae}$ in the action (\ref{S}) 
is the same as in ae-theory, as is the relation \cite{Carroll:2004ai} 
between
the cosmological gravitational constant $G_{\rm cosmo}$
that appears in the Friedmann equation, because these
relations can be obtained with hypersurface orthogonal
solutions. 
This agrees with the relations reported 
in Ref.\ \cite{Blas:2009qj}.
Similarly, the parametrized post-Newtonian (PPN) 
parameters $\b$ and $\g$ must have 
the same values as in ae-theory, in particular they 
are the same (unity) as in 
general relativity \cite{Eling:2003rd}. 
The preferred frame PPN parameters $\alpha_{1,2}$ characterize solutions 
for sources in motion with respect to the aether. These parameters might 
be different in the two theories, since the aether in such solutions 
is not necessarily hypersurface orthogonal.


\acknowledgments
I thank D. Blas, C.T.\ Eling, B.Z.\ Foster, L. Freidel, D. Mattingly, and T. Sotiriou
for helpful discussions and/or comments on the manuscript, 
and the Perimeter Institute for 
hospitality at the workshop ``Gravity at a Lifshitz Point". This research
was supported in part by the NSF under grant No. PHY-0903572. \\

\end{document}